\documentclass{article}
\usepackage[utf8]{inputenc}
\usepackage{lineno,hyperref}
\usepackage{cleveref}
\usepackage{xspace} 
\usepackage{caption}
\usepackage{graphicx}
\usepackage{here}
\usepackage{hhline}
\usepackage{comment}
\usepackage[a4paper]{geometry}
\usepackage{color}
\graphicspath{ {./images/} }
\usepackage{tikz}
\usetikzlibrary{arrows.meta, positioning, calc, fit, backgrounds, patterns}
\newcommand{\mue}{$\mu \to e$\xspace}

\begin{document}

\begin{titlepage}

\begin{center}

\vskip .55in

\begingroup
\centering


{\large\bf Extracting Signal Electron Trajectories in the COMET Phase-I Cylindrical Drift Chamber Using Deep Learning} 

\endgroup

\vskip .4in

\renewcommand{\thefootnote}{\fnsymbol{footnote}}
{
Fumihiro Kaneko$^{(a)}$\footnote{
  \href{mailto:fkaneko@preferred.jp}
  {\tt fkaneko@preferred.jp}
  },
Yoshitaka Kuno$^{(b,e)}$\footnote{
  \href{mailto:kuno@rcnp.osaka-u.ac.jp}
  {\tt kuno@rcnp.osaka-u.ac.jp}},
Joe Sato$^{(c)}$\footnote{
  \href{mailto:sato-joe-mc@ynu.ac.jp}
  {\tt sato-joe-mc@ynu.ac.jp}}, 
Ikuya Sato$^{(d)}$\footnote{
  \href{mailto:silvern198@gmail.com}
  {\tt silvern198@gmail.com}}, 
Dorian Pieters$^{(e)}$\footnote{
  \href{mailto:dopieters@gmail.com}
  {\tt dopieters@gmail.com}}, \\
 and
Chen Wu $^{(b,f,g)}$\footnote{
  \href{mailto:wuchen@ihep.ac.cn}
  {\tt wuchen@ihep.ac.cn}}
}

\vskip 0.2in

\begingroup\small
\begin{minipage}[t]{0.9\textwidth}
\centering\renewcommand{\arraystretch}{0.9}
{\it
\begin{tabular}{c@{\,}l}
$^{(a)}$
& Preferred Networks, Inc. Tokyo 100-0004, Japan \\[2mm]
$^{(b)}$
& Research Center of Nuclear Physics, Osaka University, Osaka 565--0871, Japan \\[2mm]
$^{(c)}$
& Department of Mathematics, Physics, Electrical Engineering and Computer Science,\\
&Yokohama National University, Yokohama 240--8501, Japan \\[2mm]
$^{(d)}$
& Department of Physics, Faculty of Science, Saitama University, Saitama 338--8570, Japan \\[2mm]
$^{(e)}$
& Department of Physics, Faculty of Science, Osaka University, Osaka 560--0043, Japan \\[2mm]
$^{(f)}$
& Institute of High Energy Physics, Chinese Academy of Sciences, Beijing 100049, China \\[2mm]
$^{(g)}$
& Spallation Neutron Source Science Center, Dongguan 523808,
China \\
\end{tabular}
}
\end{minipage}
\endgroup

\end{center}

\vskip .4in
    
\begin{abstract}
We present a pioneering approach to tracking analysis within the COMET Phase-I experiment, which aims to search for the charged lepton flavor violating $\mu\to e$ conversion process in a muonic atom, at J-PARC, Japan. This paper specifically introduces the extraction of signal electron trajectories in the COMET Phase-I cylindrical drift chamber (CDC) amid a high background hit rate, with more than 40\,\% occupancy of the total CDC cells, utilizing deep learning techniques of semantic segmentation. Our model achieved remarkable results, with a purity rate of 98\,\% and a retention rate of 90\,\% for CDC cells with signal hits, surpassing the design-goal performance of 90\,\% for both metrics. This study marks the initial application of deep learning to COMET tracking, paving the way for more advanced techniques in future research.
\end{abstract}

\thispagestyle{empty}
\end{titlepage}

\section{Introduction}
The COMET experiment aims to search for charged lepton flavor violation (CLFV), with a specific emphasis on neutrinoless coherent muon to electron conversion (\mue conversion) process in a muonic atom, namely,
\begin{equation}
    \mu^{-}+N(A,Z) \to e^{-}+ N(A,Z),
\end{equation}
where $N(A,Z)$ denotes a nucleus with mass number $A$ and atomic number $Z$.
This CLFV process is considered to be a promising avenue for discovering new physics beyond the Standard Model (SM)\cite{Kuno:1999jp}. 
In the minimal SM, neutrinos were thought to be massless, and lepton flavor conservation was believed to hold within each generation. However, neutrino oscillation experiments revealed that neutrinos have mass and can change flavors, disproving the assumption of strict lepton flavor conservation. However, the SM predicts an extremely small contribution to CLFV, with a rate on the order of $10^{-54}$, offering a unique experimental avenue to explore new physics beyond the SM (BSM) without significant interference from SM backgrounds. 
Expected sensitivities of the upcoming CLFV experiments, including the COMET experiment, will allow the exploration of new physics up to energy scales of O($10^4$) TeV, surpassing the capabilities of collider experiments. Thus, CLFV research serves as an important complement to Beyond the Standard Model (BSM) searches at the Large Hadron Collider.

The COMET experiment will be conducted at the Japan Proton Accelerator Research Complex (J-PARC), Tokai, Japan. COMET stands for Coherent Muon to Electron Transition.
The experiment will proceed in a staged approach, where Phase-I of the COMET experiment will aim at an experimental sensitivity of $7 \times 10^{-15}$ at a 90 \% confidence level. It is about 100 times improvement over that latest result of \cite{Bertl:2006up}. Phase-II, the second stage of the COMET experiment, will aim at further improvements in sensitivity, down to the order of $\mathcal{O}(10^{-17})$. A refinement of additional factor of ten, down to $\mathcal{O}(10^{-18}$) has been attempted. 


\begin{figure}[b!]
\begin{center}
\includegraphics[width=\textwidth]{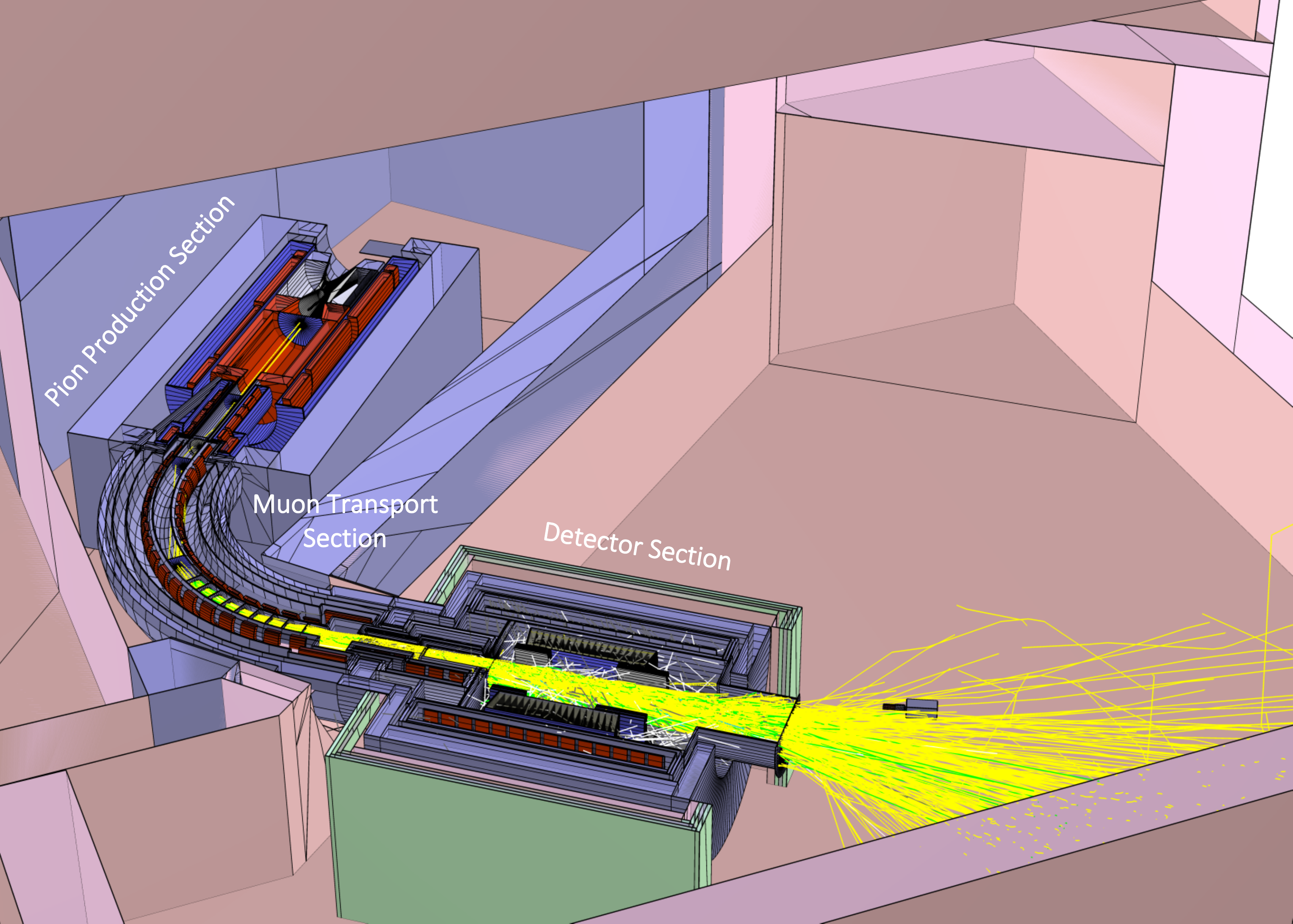}
\caption{Schematic layout of the COMET Phase-I experiment, showing the pion capture solenoid (on the left) and the 90° bend of the muon transport beam line with tunable momentum selection. The detector, consisting of the cylindrical drift chamber (CDC) and the detector solenoid, is placed at the end of the muon transport section.}
\label{fig:cometphase1}
\end{center}
\end{figure}

The schematic layout of the COMET Phase-I experiment \cite{comettdr} is given in \Cref{fig:cometphase1}.
Muons will be produced from the pions produced in the collisions of 8\,GeV protons with a proton target made of graphite.  
The yield of low momentum muons transported to the experimental area is enhanced using a superconducting 5\,T pion capture solenoid surrounding the proton target in the pion capture section in \Cref{fig:cometphase1}. 
Muons are momentum- and charge-selected using 90$^{\circ}$ curved superconducting solenoids in the muon transport section of 3\,T, before being stopped in a muon-stopping target made of aluminium located in the detector section.  

\begin{figure*}[t]
 \begin{center}
 \includegraphics[width=0.9\textwidth]{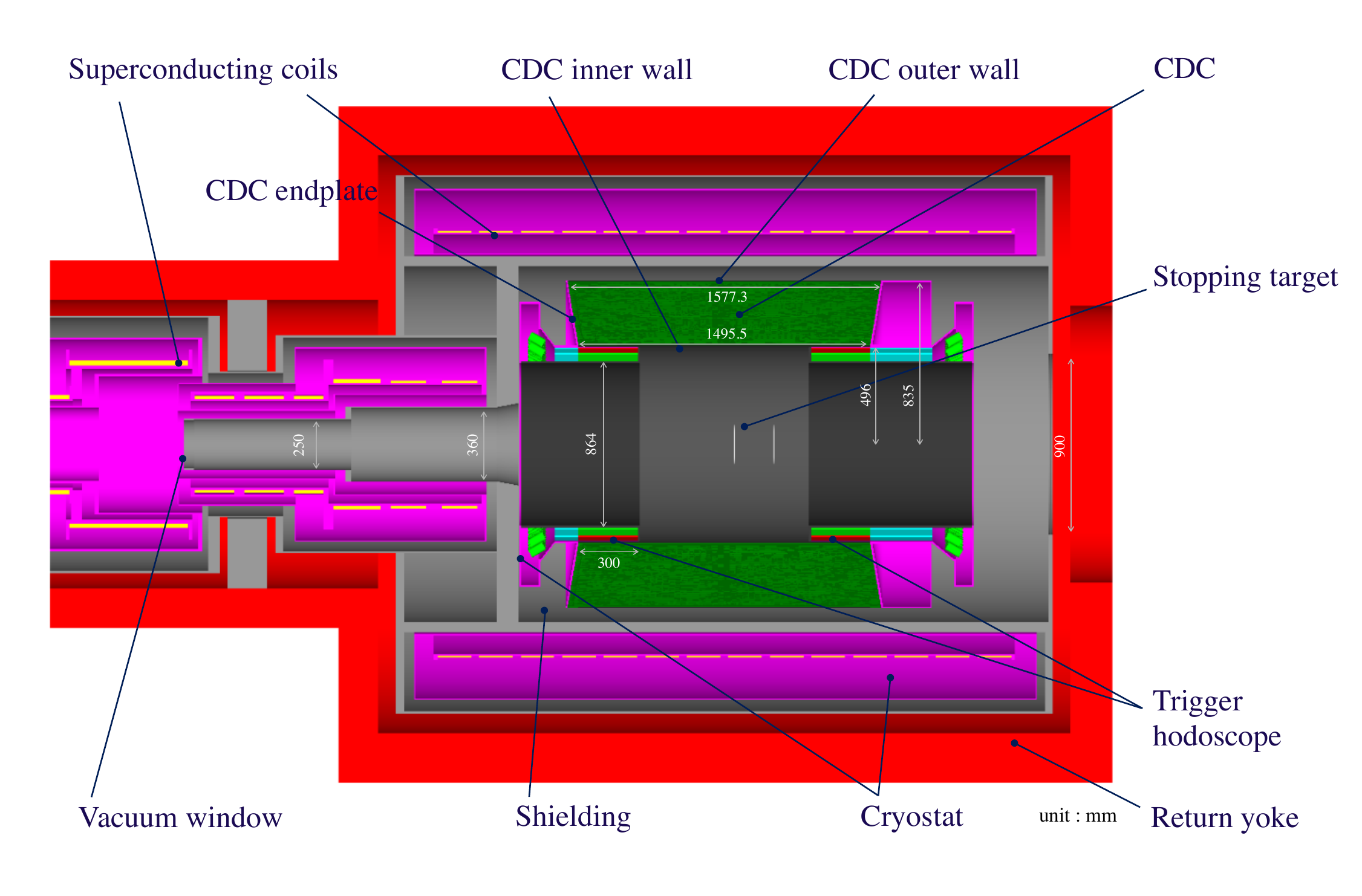}
 \end{center}
 \caption{\sl Schematic layout of the CDC detector. The upstream and downstream cylindrical trigger hodoscopes (CTH) are shown. The muon beam enters from the left and is stopped in the stopping target located inside the CDC. }
 \label{fig:cdc2d}
\end{figure*}

Figure \ref{fig:cdc2d} illustrates the COMET Phase-I detector layout, featuring the Cylindrical Drift Chamber (CDC). Being placed after the muon transport section, the CDC resides within the warm bore of a large 1\,T superconducting detector solenoid. Additionally, Cylindrical Trigger Hodoscopes (CTH), consisting of two layers of plastic scintillation counters, are placed upstream and downstream on the inner side of the CDC to trigger events and provide reference timing of the events. The CDC is essential for measuring the momentum of electrons from \mue conversion, enabling accurate identification of signal electrons and effective rejection of background events.

\section{COMET Tracking Analysis}

The physics analysis focuses on identifying signal events and evaluating background contributions. For \mue conversion, signal events are primarily identified by their momentum. In the COMET Phase-I experiment with an aluminum target, the signal momentum is 104.9\,MeV/c. Accurate momentum tracking in the CDC is therefore crucial to the success of COMET Phase-I.

A major challenge in CDC tracking lies in effectively distinguishing signal hits from numerous background hits, originating mainly from beam flashes and products resulting from muon decays and muon nuclear capture at the muon target. With the world's highest muon beam intensity employed in COMET Phase-I of $\mathcal{O}(10^9)$ stopped muons/s, being over 10 times greater than that at Paul Scherrer Institute (PSI), Switzerland, the scale of background hits is potentially substantial. 
From the studies of this issue with simulation data, an anticipated background hit occupancy of approximately 40\,\% or more for all active cells in the CDC emerges. In addition, potential fluctuations in intensity of each pulsed proton beam bunch by a factor of two or more has been identified in the past proton beam studies at J-PARC. This emphasises the critical need to assess how efficiently background hits can be rejected while retaining as many signal hits as possible. The success of CDC tracking performance depends on optimizing this balance, addressing the  complexities posed by the unprecedented intensity of muon beams in the COMET Phase-I experiment.

Previously, the exploration of signal hit extraction methods has included the use of gradient-boosted decision trees (GBDT) and the application of the Hough transformation, both of which are valuable for detecting circles amid significant background hits. These approaches have demonstrated good efficiency up to a background hit occupancy of approximately 15\,\%. \cite{comettdr}.
However, the increasing uncertainty regarding the actual background hit occupancy, coupled with potential fluctuations in proton intensity by a factor of two or more, requires a quest for a more robust and resilient method.

\begin{figure}[!htb]
 \begin{center}
 \includegraphics[width=0.6\textwidth]{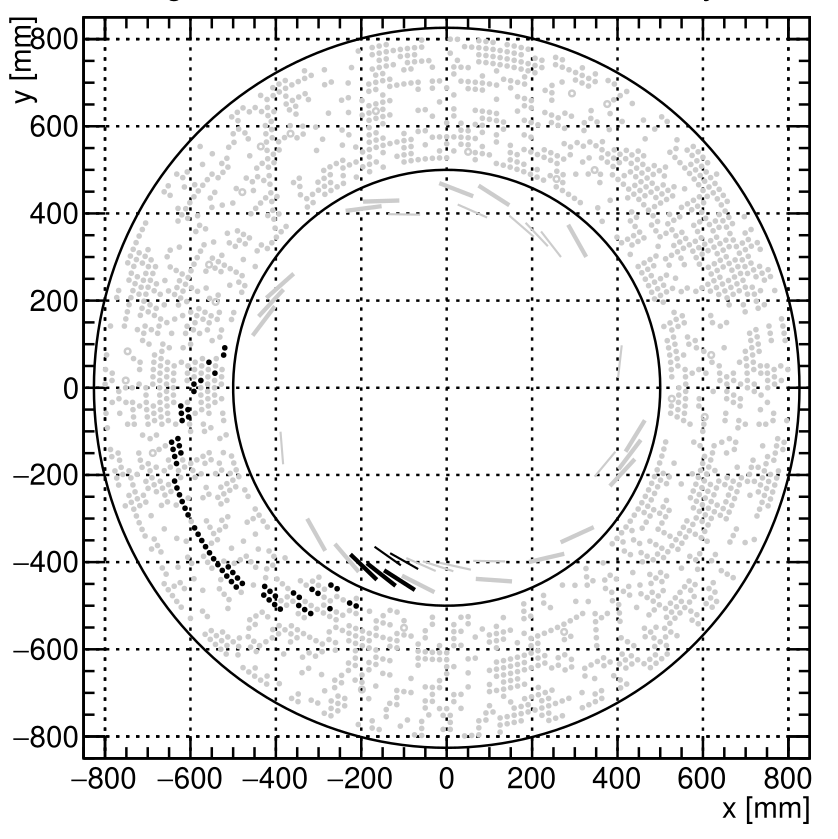}
 \end{center}
 \caption{\sl Typical event display in the CDC from simulation, with signal hits in black and background hits in gray. The level of hit occupancy for this event is 44.3\,\%.}
 \label{fig:eventdisplay}
\end{figure}

Given the expected difficult experimental conditions, it becomes crucial to develop an advanced signal hit extraction method capable of accommodating background occupancy of up to 40\,\% or more. 
Figure \ref{fig:eventdisplay} shows a typical display of hits in the CDC for one event window from simulations, with the level of hit occupancy of 44.3\,\%. At this noise level, direct track fitting of the signal track is not feasible. The first step in tracking is to separate signal hits from background hits, a process known as hit filtering or signal hit extraction, which is discussed in this paper.
The CDC contains approximately 4400 channels, with a typical signal track leaving about 60 hits in the CDC. Consequently, the ratio of background hits to signal hits is very large, necessitating an extremely effective hit filtering step. The goal of hit filtering is to provide a clean sample of signal hits with minimal background contamination while maintaining high efficiency.

For effective hit filtering,
various features of the hits are utilized. These include the spatial location of the cells with hits and specific characteristics of each hit recorded by the CDC. The CDC records timestamps of incoming pulses (hit timing) relative to the CTH reference timing and the sizes of these pulses (hit ADCs), which reflect the charge deposited by the hits. Additionally, the relative location of the cell to the CTH trigger counter is valuable information, as the track spans a limited spatial region close to the CTH trigger counter. 
This azimuthal angle information from the CTH trigger counter, referred to as $\Delta\phi$, is utilized. 

By representing this data as a 2-Dimensional (2D) image, a neural network model for semantic segmentation can efficiently 
perform hit filtering. 
For this purpose, U-Net has been selected for testing and study, and its performance will be discussed in this paper.

\restoregeometry

\section{Data Preparation}  
To prepare the data sample for this study, we began with a comprehensive Monte Carlo simulation of the muon beam. A resampling technique was employed to generate a large sample of background hits, considering the pulsed structure of the proton 
beam. Separately, a dedicated simulation of the signal electrons was conducted using the full geometry of the COMET Phase-I CDC 
detector. The sample of signal electrons was then merged with the aforementioned background hit events, accounting for overlapping effects. Finally, the merged sample was translated into images to be used as an input for the neural network.

\subsection{Beam and Detector Simulations} 
The full simulation of the muon beam begins with protons hitting the proton 
target to produce pions. 
This was carried out using the COMET standard software framework (ICEDUST \cite{comettdr}), which utilizes Geant4. The simulation included the complete geometry of the experimental hall and a comprehensive magnetic 
field map, taking fringe fields into account. In 
\Cref{fig:cometphase1} 
a display of a small fraction of the muon beam in this simulation is shown.

After simulating the muon beam, a simplified detector response simulation was performed to model the response of the working gas in the CDC, based on Garfield simulation results. In this step, waveforms from the CDC readout electronics were generated based on the incident particles recorded in the CDC from the aforementioned muon beam simulation. The waveforms were then converted into pairs of ADC and timing, 
which we refer to as hits hereafter.

The generated hits were then arranged in a time sequence according to the pulsed structure of the incident 
proton beam, which has a repetition rate of about 1\,MHz.

Assuming that the intensity of the proton pulse remains constant during the physics run, each incident proton pulse contains 16 million protons hitting the target. The simulated data sample was thus converted into 300 pulses. Unfortunately, due to the presence of very slow processes like neutron propagation, the sequence of hits from 300 pulses cannot accurately reflect the radiation level after beam-on. To overcome this limitation, we adopted a conservative approach by adding a 300-microsecond modulo to the hit time. This transforms the time sequence of hits into a ring structure, effectively simulating the scenario of an infinite number of preceding proton beam pulses.

\begin{figure}[!htb]
 \begin{center}
 \includegraphics[width=0.6\textwidth]{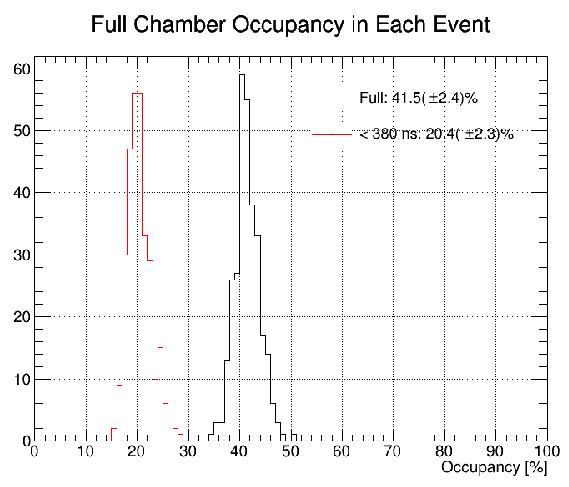}
 \end{center}
 \caption{\sl The distribution of CDC hit occupancy by background hits from the beam simulation.}
 \label{fig:occupancy}
\end{figure}

\begin{figure}[!htb]
 \begin{center}
 \includegraphics[width=0.6\textwidth]{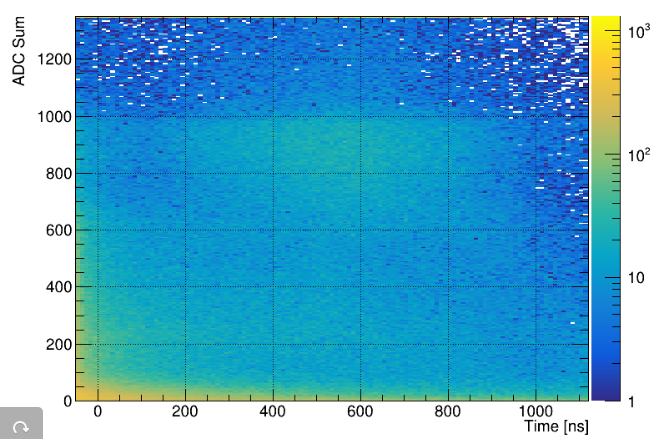}
 \end{center}
 \caption{\sl The distribution of timing in horizontal and charge (in ADC unit) in vertical for background hits from the beam simulation.}
 \label{fig:noise2d}
\end{figure}

\begin{figure}[!htb]
 \begin{center}
 \includegraphics[width=0.6\textwidth]{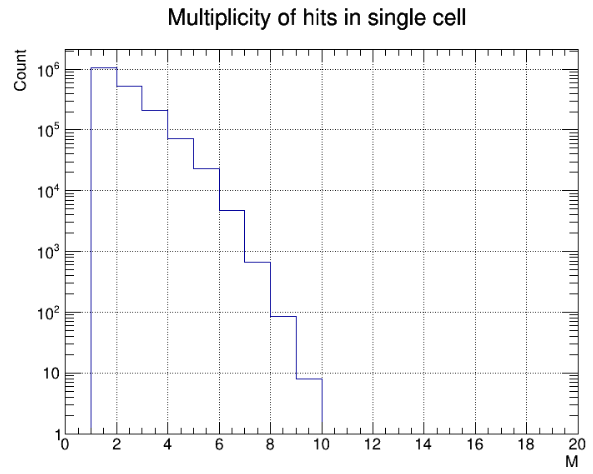}
 \end{center}
 \caption{\sl The distributions of multiplicity of hits in each occupied cell in one event window, from the 
 simulation. The average multiplicity is $1.7$.}
 \label{fig:noisemultiplicity}
\end{figure}

Eventually, the hits sequence ring was divided into 300 event frames to examine the background hits distribution. 
The average hit occupancy across the full time range is approximately 40\,\%, as shown in \Cref{fig:occupancy}. 
The 2-dimensional (2D) distribution of charge and timing for all the background 
hits are shown in \Cref{fig:noise2d}.
It should be noted that there are sometimes multiple hits in one cell from a single event. The multiplicity distribution according to this simulation is shown in \Cref{fig:noisemultiplicity}

\subsection{Background Hit 
Samples from Resampling} \label{sec: bkg_resampling}
Due to the limited computing resources, it is not practical to directly generate millions of events for our study using the full simulation method. Instead, we employed a resampling method based on the distributions mentioned above to create a sufficient number of background hit samples.
It is important to note that this method cannot account for the correlation among individual hits. As a result, some cluster patterns of the background hits in the CDC, which are observable from the full beam simulation, are not preserved in the samples generated using the resampling method. The performance, considering cluster patterns, will be studied in future work.

\subsection{Signal Electron Samples 
}
The signal electrons are mono-energetic with a momentum of 104.9\,MeV/$c$ for the case of a muonic atom with aluminum.
We 
used ICEDUST to simulate the signal electrons originating 
from the muon stopping target. The starting vertices of the signal electrons were resampled from the distribution of positions and timing of muons stopped in the muon stopping target, using the muon beam simulation. 
An additional time delay, due to the lifetime of muonic atom of aluminum (864\,ns), 
was added to the timing of each signal electron.

After the simulation, an appropriate detector response simulation was performed to generate hits in the CDC. 
Additionally, hits in the CTH were also created to examine the trigger conditions. Using the hits recorded in each event, we developed an algorithm to select events that met 
the trigger requirement in the CTH (the 4-fold coincidence) and the geometrical requirement in the CDC, which demands that the signal track should pass through at least 5 layers to provide sufficient information on the longitudinal direction.

\begin{figure}[!htb]
 \begin{center}
 \includegraphics[width=0.6\textwidth]{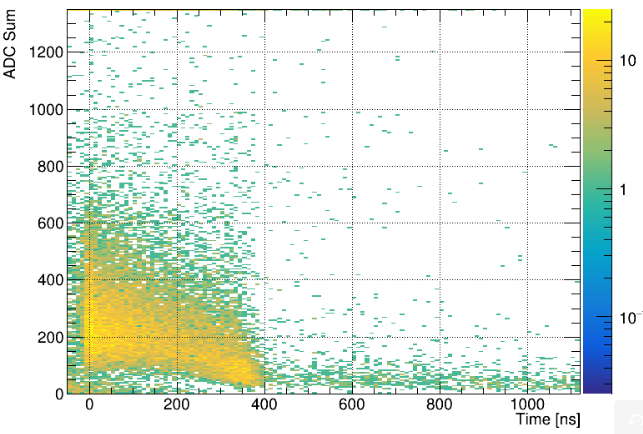}
 \end{center}
 \caption{\sl The distribution of timing in horizontal and charge (in ADC unit) in vertical for signal hits, from the signal electron simulation.}
 \label{fig:signal2d}
\end{figure}

After the selection, a pure signal sample was generated. The typical number of signal hits per signal electron is about 66. The 2D distribution of timing versus charge for all signal hits is shown in Figure \ref{fig:signal2d}.

\subsection{Merging Signal and Background Samples}
The pure signal sample was then merged with the previously generated background hits sample. To account for the overlap effect between signal and background events, we applied a simple treatment assuming a fixed signal width of 100\,ns. This overlap treatment resulted in the loss of approximately 20\,\% of the signal hits.

The final merged sample was then used to create images for input into the neural network. These images included various distributions and patterns that the neural network would need to learn and recognize, thereby ensuring that the network could be effectively trained on realistic data representative of actual experimental conditions.

\subsection{Creation of Images}

To create images for the neural network, hits are converted to pixels by assigning hit attributes to pixel colors and arranging the pixels onto a canvas. These pixels are then grouped by signal or background labels.

\subsubsection{Cells with Multiple Hits}

To ensure all images have the same canvas size, hits are arranged on the canvas according to the cells they originate from. When a cell has multiple hits, the corresponding pixel needs to represent features from multiple hits. \Cref{fig:noisemultiplicity} shows the distribution of hit multiplicity per cell according to full simulation data. This issue can be addressed by adding additional color channels to each pixel of the image. However, at this stage of the study, we aim to keep the pixel features simple to maintain the robustness of the algorithm. Therefore, we decided to select one hit to represent the feature of a cell, and then assign these features as colors to the corresponding pixel. Pixels corresponding to a cell containing a signal hit are labeled as signal, while the rest are labeled as noise. The algorithm's task is to identify the signal pixels, and hence, the cells with signal hits. In cases where a cell has multiple hits, the subsequent tracking procedure will determine which hit from the selected cells to use in the track fitting step.

This approach still requires using one hit to determine the features of a cell. We employed two methods to select which hit to use. The first method uses the hit timing and hit ADC distribution of signal hits, as shown in \Cref{fig:signal2d}. 
In this method, a 2D distribution is created with each axis representing a distinct feature: hit timing and ADC value. This distribution is populated with signal hits from the simulation, creating a density map. The density at each point in this 2D distribution 
indicates the likelihood that a hit with those particular time and ADC values is a signal hit. For a cell with multiple hits, we examine the time and ADC values of each hit and select the hit that corresponds to the highest density (or likelihood) in the 2D distribution. By selecting the hit with the highest likelihood, we can achieve reasonable prediction accuracy.

For comparison, we also tested a simpler method that selects the first hit in each cell. In the initial evaluation, both methods demonstrated roughly comparable performance, likely due to the low hit multiplicity in each cell. A more detailed study of comparison will follow in this paper.

\subsubsection{Converting Hits to Pixels} \label{sec:hits_to_pixs}

After reducing the dimensionality by selecting only one hit per cell, each pixel corresponds to a single hit. Thus, only three color channels are needed for pixels. The first channel represents the hit timing, the second channel represents the ADC value, and the third channel represents $\Delta\phi$, the azimuthal angle between the CTH trigger counter and each cell.

\begin{figure}[!htb]
    \centering
    \includegraphics[width=1\linewidth]{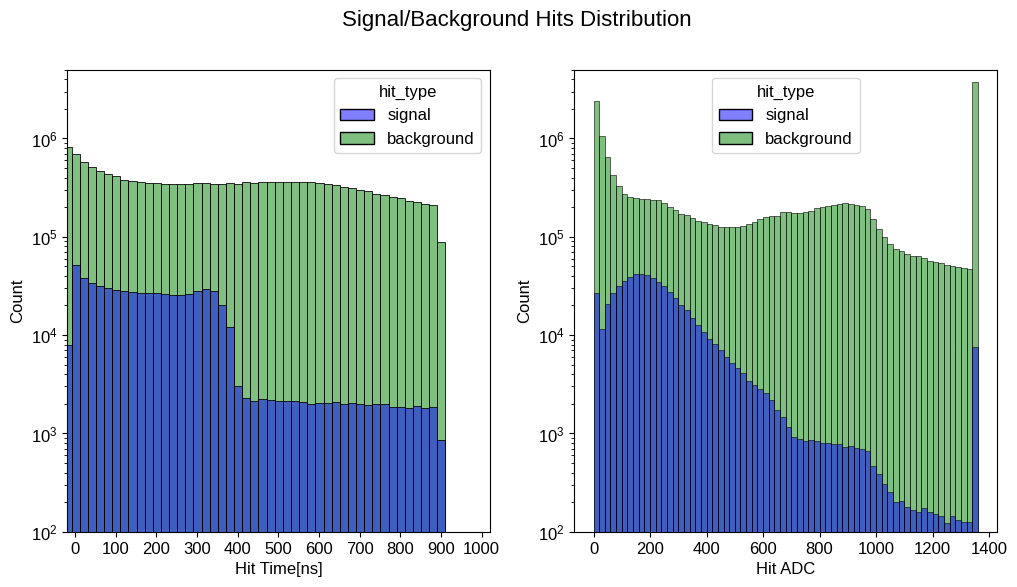}
    \caption{Signal and background hit distributions for hit timing (left) and hit ADC  (right) are shown with logarithmic-scaled y-axis. Signal hits cluster within specific ranges, while background hits are more uniformly distributed. 
    }
    \label{fig:signal_bkg_distribution}
\end{figure}

The three types of input values can be converted to color values using simple linear scaling. However, one feature in the distribution of signal hits suggests that scaling with a limited input value range might be beneficial for maintaining precision with a limited number of bits for color values. In the simulation, as shown in \Cref{fig:signal_bkg_distribution}, 
most signal 
electron hits fall within the range ($-$50.0 ns, 
400 ns) 
for time information and (0, 700) for the ADC information. In the limited scaling scheme, hits within these ranges are linearly scaled, while those outside are assigned capped values.

When mapping hit cells to pixels, some pixels correspond to non-existing 
cells, while some correspond to cells with no hits. To distinguish these special pixels from others, black is reserved for pixels without corresponding cells, while white is used for pixels representing empty cells. Narrow gaps in the color value range are maintained around both black and white to ensure these special cases are clearly distinguishable from regular hit data. 

\subsubsection{Arranging Pixels on Canvas}

In our study, we utilize two distinct image formats derived from the detector cell configuration to train the neural network: the "normal" format and the "reduced" format.

\begin{figure}[!htb]
\begin{center}
  \begin{minipage}[b]{0.45\linewidth}
    \centering
    \includegraphics[width=50mm]{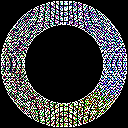}
    \caption*{(a) input image with normal format, 128$\times$128}
  \end{minipage}
  \begin{minipage}[b]{0.45\linewidth}
    \centering
    \includegraphics[width=50mm]{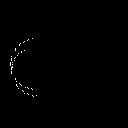}
    \caption*{(b) label with normal format, 128$\times$128}
  \end{minipage}
  \begin{minipage}[b]{0.45\linewidth}
    \centering
    \includegraphics[width=50mm]{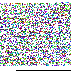}
    \caption*{(c) input image with reduced format, 71$\times$71}
  \end{minipage}
  \begin{minipage}[b]{0.45\linewidth}
    \centering
    \includegraphics[width=50mm]{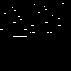}
    \caption*{(d) label with reduced format, 71$\times$71}
  \end{minipage}

\caption{(a),(b) The input image and label of the normal format. (c),(d) The input image and label of the reduced format. Each input image has 3 channels: red for hit timing, green for ADC of hits, and blue for $\Delta\phi$ of hits. White pixels of input image indicate cells with no hits, while black pixels represent areas not corresponding to any cells in the CDC detector. The label images are gray scale, with white pixels denoting cells with true signal hits.}
\label{figure:input-output}
\end{center}
\end{figure}

\paragraph{Normal Format (for \Cref{figure:input-output}(a)(b))}
In the "normal" format, the detector’s 2-dimensional plane—more specifically, a projection of cells at the CDC 
readout plane—is treated as an image, using (x, y) coordinates to define the plane on the CDC 
section. The cell size is approximately 16.8 mm wide and 16.0 mm high, with the radius of the CDC 
readout at 818 mm \cite{comettdr}. To represent each detector cell as a single pixel, the plane is discretized into 128 bins along both the x and y axes, resulting in a 128$\times$128 pixel image. This results in a grid size of 12.8 mm per bin (1636 mm / 128), which is smaller than the actual cell size. Although the cells are arranged along the section of a cylinder, requiring a finer grid when using Cartesian coordinates, this discretization is essential for ensuring each cell is accurately represented by a single pixel. However, it is important to note that approximately 70\% of this image area remains unoccupied(black area of Figure \ref{figure:input-output}(a)), rendering this format inefficient for neural network training.

\paragraph{Reduced Format (for \Cref{figure:input-output}(c)(d))}
To address the inefficiency of the normal format, we introduced the "reduced" format. Here, cells are indexed sequentially from the inner to the outer regions, yielding indices from 0 to 4987. The cells are then arranged in the image from top-left to bottom-right, creating a compact 71$\times$71 image. Although this format is less intuitive visually, as it abstracts the spatial relationships between cells, it significantly reduces the proportion of empty space, thus improving the training efficiency of the neural network.

\paragraph{Comparison of Formats}
The both formats were evaluated in our study to determine their effectiveness in neural network training. The normal format offers a spatially intuitive representation, which may benefit visualization and interpretability. In contrast, the reduced format, despite its abstract representation, offers more efficient pixel space utilization, leading to potentially faster training. The results of this comparison are discussed in the following sections.

\section{Deep Learning Model for Tracking in the COMET Experiment}
\subsection{Introduction to Deep Learning Model} 

When considering the event as an image, the task of signal track extraction 
in our study can be considered 
as a semantic segmentation task within the realm of image recognition. Generally, semantic segmentation involves classifying each pixel in an image into distinct categories. In our study, we adapt this concept to classify each element in the image of the CDC 
readout cell into two distinct categories: signal or background. The dominant approach in semantic segmentation leverages neural networks, especially deep convolutional models, known for their robust and strong performance in complex image recognition tasks \cite{he2016deep, Redmon_2016_CVPR, He_2017_ICCV}, suggesting their promising applicability to our tracking task.  This notion is supported by previous studies \cite{MicroBooNE:2018kka, Ostdiek:2020cqz} that utilize deep convolutional models for image segmentation in the analysis of high-energy physics experiments.

In this context, we focused our investigation on a specific deep learning architecture that is well-suited for our task: the U-Net \cite{Ronneberger2015}, which is a very common architecture for semantic segmentation and is widely used in other domains including denoising in recent generative diffusion models \cite{Ho2020, Rombach_2022_CVPR}. Its simplicity and effectiveness make it an excellent baseline for our initial study on the COMET signal track extraction.

\subsubsection{U-Net}

The U-Net model, initially proposed by \cite{Ronneberger2015} for biomedical image segmentation, is designed to yield high accuracy in segmentation tasks through its symmetric encoder-decoder structure (in \Cref{figure:unet}(a)). The encoder, comprising several convolutional and max-pooling layers, effectively captures the contextual information of the input image by progressively downsampling it to extract hierarchical features. Conversely, the decoder consists of decoder blocks with skip connections from the encoder features that upsample the feature maps to the original resolution, utilizing the fine-grained feature maps from the encoder to facilitate precise localization of the segmented regions.

Although a larger model is typically better in deep learning, the original U-Net model, which employs 2 convolutional layers per stage in the encoder, has a relatively small number of parameters and layers. The method of scaling the number of parameters is not unique and sometimes requires additional architectural study to achieve the expected scaling behavior, which takes considerable time.

To address this, we opted to replace the original encoder part of U-Net with a more modern and deeper backbone model as illustrated in 
\Cref{figure:unet}(b). Introducing a backbone is common in computer vision and can be seen in recent works \cite{carion2020end, cheng2021masked}. In this study, we chose EfficientNet \cite{tan2019efficientnet} as the backbone due to its efficiency and high performance on the ImageNet benchmark \cite{ILSVRC15}, as well as its active use in machine learning competitions over the past few years \cite{compe-steel,compe-uwmg,compe-contrails} with the famous public U-Net implementation \cite{Iakubovskii:2019}. The following discussion, we call this custom U-Net as "U-Net" for simplicity.

EfficientNet, developed through neural architecture searches \cite{zoph2017neural, tan2019mnasnet}, optimizes both accuracy and efficiency by scaling up depth, width, and resolution in a balanced manner. This model builds on the success of earlier architectures like ResNet \cite{he2016deep}, incorporating advancements that allow it to achieve superior performance with fewer parameters and reduced computational cost. EfficientNet comes in a family of models ranging from EfficientNet-B0 to EfficientNet-B7, each with an increasing number of parameters and layers. This scalability allows us to evaluate different model capacities by simply changing the model index, facilitating a more systematic investigation of the model's performance on our task.

\begin{figure}[H]
    \centering
    \begin{minipage}[b]{0.495\textwidth}
        \centering
        \scalebox{0.60}{
            \begin{tikzpicture}[
                node distance=0.65cm,
                layer/.style={
                    rectangle,
                    draw=black, very thick,
                    minimum height=0.6cm,
                    minimum width=4cm,
                    align=center
                },
                label/.style={
                    align=right
                },
                arrow/.style={
                    thick,
                    ->,
                    >=stealth
                },
                connection/.style={
                    thick,
                    ->,
                    color=black
                }
            ]

            \node[layer] (input1) { };
            \node[label, fill=white, left=0.1cm of input1, font=\large\bfseries] {Input};

            \node[layer, fill=white, minimum width=4cm, below=of input1] (c0) {2$\times$Conv};
            \node[label, fill=white, left=0.1cm of c0] {C0};

            \node[layer, fill=white, minimum width=3cm, below=of c0] (c1) {2$\times$Conv};
            \node[label, fill=white, left=0.1cm of c1] {C1};

            \node[layer, fill=white, minimum width=2.5cm, below=of c1] (c2) {2$\times$Conv};
            \node[label, fill=white, left=0.1cm of c2] {C2};

            \node[layer, fill=white, minimum width=2cm, below=of c2] (c3) {2$\times$Conv};
            \node[label, fill=white, left=0.1cm of c3] {C3};

            \node[below=1.0cm of c3] (p4_y) {};
            \node[layer, fill=white, minimum width=1.5cm, right=1.8cm of p4_y] (c4_p4) {2$\times$Conv};
            \node[label, fill=white, below=0.1cm of c4_p4, align=center] {C4/P4};
            \node[right=1.8cm of c4_p4] (p4_x) {};

            \node[layer, fill=red!10, minimum width=2cm, above=1.0cm of p4_x] (p3) {Dec. Block};
            \node[label, fill=white, right=0.1cm of p3] {P3};

            \node[layer, fill=red!10, minimum width=2.5cm, above=of p3] (p2) {Dec. Block};
            \node[label, fill=white, right=0.1cm of p2] {P2};

            \node[layer, fill=red!10, minimum width=3cm, above=of p2] (p1) {Dec. Block};
            \node[label, fill=white, right=0.1cm of p1] {P1};

            \node[layer, fill=red!10, minimum width=4cm, above=of p1] (p0) {Dec. Block};
            \node[label, fill=white, right=0.1cm of p0] {P0};

            \node[layer, above=of p0] (conv_1x1) {Conv 1$\times$1};
            \node[layer, above=of conv_1x1, font=\bfseries] (seg_map) {Segmentation Map};

            \draw[arrow] (input1) -- (c0);
            \draw[arrow, dashed, very thick] (c0) -- (c1);
            \draw[arrow, dashed, very thick] (c1) -- (c2);
            \draw[arrow, dashed, very thick] (c2) -- (c3);
            \draw[arrow, dashed, very thick] (c3) -- (c4_p4);

            \draw[arrow, dashed, very thick]  (c4_p4) -- node[midway, right] {} (p3);
            \draw[arrow, dashed, very thick]  (p3) -- node[midway, right] {} (p2);
            \draw[arrow, dashed, very thick]  (p2) -- node[midway, right] {} (p1);
            \draw[arrow, dashed, very thick]  (p1) -- node[midway, right] {} (p0);
            \draw[arrow]  (p0) -- node[midway, right] {} (conv_1x1);
            \draw[arrow]  (conv_1x1) -- node[midway, right] {$\#$ of classes} (seg_map);

            \draw[arrow, very thick, blue] (c3.east) -- (p3.west);
            \draw[arrow, very thick, blue] (c2.east) -- (p2.west);
            \draw[arrow, very thick, blue] (c1.east) -- (p1.west);
            \draw[arrow, very thick, blue] (c0.east) -- (p0.west);

            \node[right=0.4cm of p4_x] (dec_block_x) {};
            \node[layer, fill=white, minimum width=3cm, below=1.0cm of dec_block_x] (detail_p) {2$\times$Conv};
            \node[layer, fill=white, minimum width=3cm, below=0.7cm of detail_p] (concat) {Channel Concatenation};

            \begin{pgfonlayer}{background}
                \node[draw=black, very thick, dashed, inner sep=15pt, pattern=crosshatch, pattern color=green!30, fit=(c0) (c1) (c2) (c3)](encoder) {};
                \node[draw=black, very thick, dashed, inner sep=15pt, fill=blue!10, fill opacity=1.0, fit=(p0) (p1) (p2) (p3)](decoder) {};
                \node[draw=black, very thick, dashed, inner sep=15pt, fill=red!10, fill opacity=1.0, fit=(detail_p) (concat) ](decoder_block) {};
            \end{pgfonlayer}{background}

            \node[below=0.1cm of encoder, font=\large\bfseries] {Encoder};
            \node[below=0.1cm of decoder, font=\large\bfseries] {Decoder};
            \node[left=0.0cm of decoder_block, font=\large\bfseries] {Decoder Block};

            \draw[arrow]  (concat)  -- (detail_p);

            \node[left=2.3cm of concat] (encoder_skip_start) {};
            \node[below=0.8cm of concat] (decoder_prev_start) {from previous decoder block};
            \node[above=0.8cm of detail_p] (decoder_next_start) {};
            \draw[arrow, very thick, blue] (encoder_skip_start) --  node[midway, below] {from encoder} (concat.west);
            \draw[arrow] (decoder_prev_start) -- (concat.south);
            \draw[arrow] (detail_p.north) --  (decoder_next_start);

            \end{tikzpicture}
        }
        \caption*{(a) Architecture of original U-Net \cite{Ronneberger2015}}
    \end{minipage}
    \begin{minipage}[b]{0.495\textwidth}
        \centering
        \scalebox{0.60}{
            \begin{tikzpicture}[
                node distance=0.8,
                layer/.style={
                    rectangle,
                    draw=black, very thick,
                    minimum height=0.6cm,
                    minimum width=4.0cm,
                    align=center
                },
                label/.style={
                    align=right
                },
                arrow/.style={
                    thick,
                    ->,
                    >=stealth
                },
                backbone/.style={
                    rectangle,
                    draw=black, very thick,
                    minimum height=4.7cm,
                    minimum width=3.5cm,
                    align=center,
                    fill=orange!30
                }
            ]

            \node[layer] (input1) { };
            \node[label, fill=white, left=0.1cm of input1, font=\large\bfseries] {Input};

            \node[backbone, below=of input1, anchor=north] (backbone) {Backbone};

            \node[below=1.0cm of backbone.south] (p5_y) {};
            \node[layer, fill=white, minimum width=1.5cm, right=1.7cm of p5_y] (c5_p5) {Identity};
            \node[label, fill=white, below=0.1cm of c5_p5, align=center] {C5/P5};
            \node[right=1.7cm of c5_p5] (p5_x) {};


            \node[layer, fill=red!10, minimum width=2.0cm, above=1cm of p5_x] (p4) {Dec. Block};
            \node[label, fill=white, right=0.1cm of p4] {P4};

            \node[layer, fill=red!10, minimum width=2.5cm, above=of p4] (p3) {Dec. Block};
            \node[label, fill=white, right=0.1cm of p3] {P3};

            \node[layer, fill=red!10, minimum width=3.0cm, above=of p3] (p2) {Dec. Block};
            \node[label, fill=white, right=0.1cm of p2] {P2};

            \node[layer, fill=red!10, minimum width=3.5cm, above=of p2] (p1) {Dec. Block};
            \node[label, fill=white, right=0.1cm of p1] {P1};

            \node[layer, fill=red!10, minimum width=4.0cm, above=of p1] (p0) {Dec. Block};
            \node[label, fill=white, right=0.1cm of p0] {P0};

            \node[layer, above=of p0] (conv_1x1) {Conv 3$\times$3};
            \node[layer, above=of conv_1x1, font=\bfseries] (seg_map) {Segmentation Map};

            \draw[arrow] (input1) -- (backbone);


            \draw[arrow, dashed, very thick] (c5_p5) -- node[midway, right] {} (p4.south);
            \draw[arrow, dashed, very thick] (p4) -- node[midway, right] {} (p3);
            \draw[arrow, dashed, very thick] (p3) -- node[midway, right] {} (p2);
            \draw[arrow, dashed, very thick] (p2) -- node[midway, right] {} (p1);
            \draw[arrow, dashed, very thick] (p1) -- node[midway, right] {} (p0);
            \draw[arrow] (p0) -- node[midway, right] {} (conv_1x1);
            \draw[arrow]  (conv_1x1) -- node[midway, right] {$\#$ of classes} (seg_map);

            \draw[arrow] (backbone.south) -- (c5_p5);
            \draw[arrow, very thick, blue] (backbone.east)++(0,-1.8cm) -- ++(0cm,0cm) |- (p4.west);
            \draw[arrow, very thick, blue] (backbone.east)++(0,-0.4cm) -- ++(0cm,0cm) |- (p3.west);
            \draw[arrow, very thick, blue] (backbone.east)++(0,1.0cm) -- ++(0cm,0cm) |- (p2.west);
            \draw[arrow, very thick, blue] (backbone.east)++(0,2.1cm) -- ++(0cm,0cm) |- (p1.west);

            \begin{pgfonlayer}{background}
                \node[draw=black, very thick, dashed, inner sep=15pt, pattern=crosshatch, pattern color=green!30, fit=(backbone)](encoder) {};
                \node[draw=black, very thick, dashed, inner sep=15pt, fill=blue!10, fill opacity=1.0, fit=(p0) (p1) (p2) (p3) (p4)](decoder) {};
            \end{pgfonlayer}

            \node[below=0.1cm of encoder, font=\large\bfseries] {Encoder};
             \node[below=0.1cm of decoder, font=\large\bfseries] {Decoder};

            \end{tikzpicture}
        }
        \caption*{(b) Architecture of U-Net with replacing encoder part with other backbone model \cite{Iakubovskii:2019}}
    \end{minipage}
    \caption{Architecture of U-Net. (a) The original U-Net encoder comprises several convolutional blocks and downsamplings, producing a series of spatial compressed feature maps \{C0, C1, C2, C3, C4\}, corresponding to different sizes \{1, $1/2$, $1/4$, $1/8$, $1/16$\} relative to the input image. "2$\times$Conv" indicates a sequence of two 3$\times$3 convolution layers each followed by a ReLU activation. The dashed upward and downward arrows indicate upsampling and downsampling (max-pooling) operations with a factor of 2, respectively. (b) In the custom U-Net, the encoder is replaced with a backbone model which is typically optimized for famous image recognition benchmarks like ImageNet classification. The backbone also produces different feature maps (e.g., C1-C5) similar to the original U-Net encoder. Thus, the custom U-Net can utilize larger models that are well-studied in their architecture on other benchmarks. The decoder in both architectures (a) and (b) refines these features by sequentially upscaling them with the decoder block using skip connections between each upscaled feature map, P0-P4, and its corresponding encoder feature map, C0-C4. The final layer of the U-Net uses a convolution layer to map the output to the desired number of classes. In the custom U-Net, we follow the famous public implementation \cite{Iakubovskii:2019} and modify the original architecture, like inserting batch normalization layers into the decoder block and using 3$\times$3 convolutions in the final layer.}
    \label{figure:unet}

\end{figure}

For our initial research into the COMET signal track extraction, we employed the U-Net architecture with EfficientNet backbone to harness its advanced capabilities without extensive hyperparameter customization.

\section{Implementation Details}
In this section, we delve into the details of training deep learning models tailored for semantic segmentation in the image of the CDC 
readout cells from synthetic events in the COMET experiments.

Our dataset comprises 90,000 synthetic events, divided into training, validation, and testing sets with an 80:10:10 distribution ratio. This division is stratified based on the number of signals to ensure equal distribution across these sets.

By default, the input image is set to 288$\times$288 for its size and the reduced format. However, to explore the dependencies of the model on input size and format, we conducted several training experiments with varying input sizes, ranging from 128$\times$128 to 842$\times$842, and both the normal and reduced formats. Nearest interpolation is used to resize the input images to preserve hit information accurately. The target maps retain their original sizes (71$\times$71 for reduced format or 128$\times$128 for normal format), and the model outputs are downsampled to these original sizes to maintain the classification target with linear interpolation. This setup defines the segmentation task as a pixel-wise classification into signal (1) or background (0) classes, using binary cross-entropy loss as the learning criterion. When evaluating the loss function, we ignore the loss values from the unoccupied area in the input image (black area of Figure \ref{figure:input-output}(a)(c)).

Our preprocessing methods include options such as linear and limited scaling for the scaling method, and the first hit and 2D distribution  for handling multi-hit events. By default, we use linear scaling for the scaling method and the 2D distribution  method for multi-hit handling.

For the model architecture, we employ EfficientNet models as U-Net backbone, scaling from EfficientNet-b0 to EfficientNet-b7 to explore the impact of parameter scaling on performance. The encoder weights are pre-trained on ImageNet to leverage the benefits of transfer learning, improving the initial performance of our models. By default, EfficientNet-b2 is used as the primary architecture.

During the validation phase, we use the Area Under Curve (AUC) for the Purity-Retention (precision-recall) curve as our main performance indicator. The model with the highest AUC validation score across training epochs is selected for optimizing the Purity-Retention curve.

Our computational infrastructure includes a PyTorch framework \cite{paszke2019pytorch} running on an RTX 4090 GPU with 24GB of memory. The hyperparameters settings are referred to \cite{liu2022convnext} and pick up major settings with minimal adaptation for our task. Further hyperparameter search could yield performance enhancements. The hyperparameters are set as follows: the batch size is 96, exponential moving average for model weights is used with a decay rate of 0.996, the AdamW optimizer \cite{adamw} is employed, a Cosine Annealing scheduler with a 10\,\% duration warmup and a peak learning rate of $3\times10^{-3}$ 
is utilized, the training process spans 30 epochs, and weight decay is set at $1.0\times10^{-3}$. 

\section{Results and Analysis}
To evaluate the final performance of the models, we employed metrics derived from "Purity-Retention" curve on the test data. This metric, based on True Positives (TP), False Negatives (FN), True Negatives (TN), and False Positives (FP), include:
\begin{itemize}
    \item \textbf{Purity} \(\frac{TP}{TP + FP}\): This metric represents the accuracy of positive predictions. This is  called precision or positive predictive value too.
    \item \textbf{Retention} \(\frac{TP}{TP + FN}\): This metric measures the proportion of correctly identified positive instances. This is called recall or true positive rate too.
    
\end{itemize}

This metric provide a comprehensive evaluation of the model's ability to accurately identify positive and negative cases, a critical aspect for addressing the requirements of COMET tracking task. And it is said that to obtain the necessary momentum resolution in COMET, this Purity-Retention curve must pass outside (0.9,0.9), so we will discuss whether such a learning result can be obtained.

\subsection{Model Performance on Signal Detection}

\begin{figure}[H]
    \includegraphics[width=1.0\linewidth]{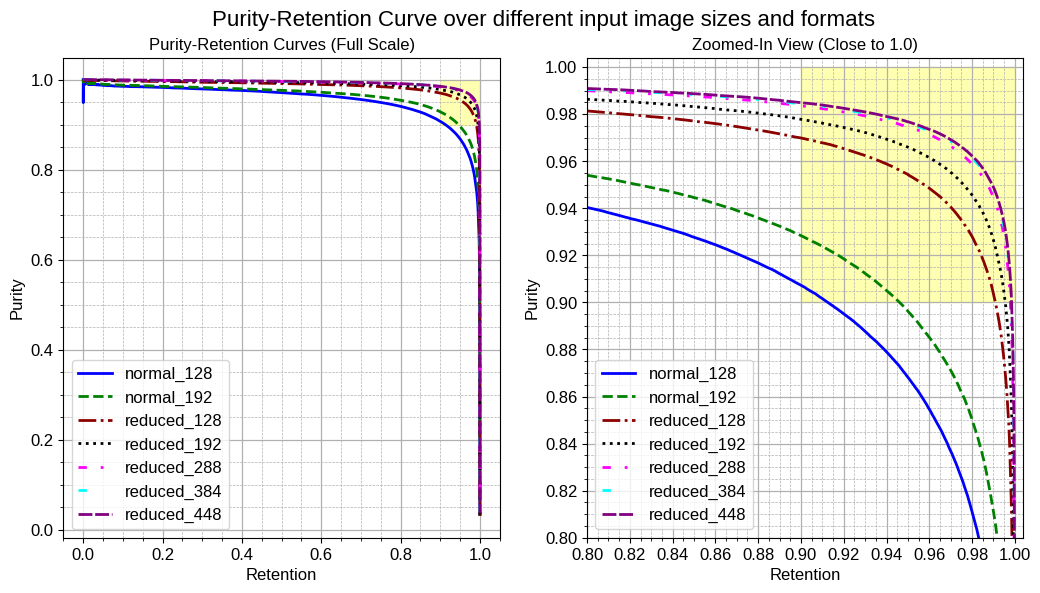}
    \caption{Purity-Retention curves for different input image sizes and formats. The left plot shows the full-scale view of the curves for various input image sizes and formats, where \textit{normal\_xxx} indicates the normal format with input size \textit{xxx}, and \textit{reduced\_xxx} indicates the reduced format with input size \textit{xxx}. The right plot zooms in on the high-purity, high-retention region (yellow area), highlighting where both metrics exceed 0.9. The reduced format generally proves more effective than the normal format, with the input image size significantly influencing the curves. Notably, several reduced formats achieve the target region of (0.9, 0.9), indicating superior performance.}
    \label{fig:pr_curve_input_image_sizes}
\end{figure}

\begin{figure}[ht]
    \centering
    \begin{minipage}[b]{0.45\textwidth}
        \centering
        \includegraphics[width=\textwidth]{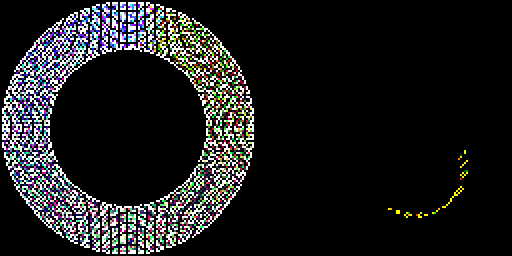}
        \caption*{(a) normal format.}
    \end{minipage}
    \hfill
    \begin{minipage}[b]{0.45\textwidth}
        \centering
        \includegraphics[width=\textwidth]{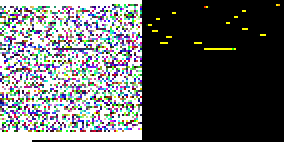}
        \caption*{(b) reduced format.}
    \end{minipage}
    \caption{Visualization of model predictions for different image formats with the input image size of 288. The left side of each image shows the input to the model, while the right side of each image shows the merged ground truth and prediction. In the right side image, ground truth is represented in red intensity, prediction in green intensity, so bright yellow indicates correctly predicted pixels.}
    \label{fig:pred_visualization}
\end{figure}

Our model achieved the target score, surpassing both purity and retention metrics of 0.9 simultaneously, as illustrated in Figure \ref{fig:pr_curve_input_image_sizes}. We can visually confirm our model successfully picks up target signal information from noisy input as shown in Figure \ref{fig:pred_visualization}. From Figure \ref{fig:pr_curve_input_image_sizes}, We identified two key factors essential for achieving the target score:

\begin{enumerate}
    \item The adoption of a reduced format, which unexpectedly outperformed the normal format.
    \item The optimization of input image size, which played a crucial role in enhancing model performance.
\end{enumerate}

Figure \ref{fig:pr_curve_input_image_sizes} presents the Purity-Retention curve for different input image sizes and formats. The left plot shows the overall curves for the \textit{normal} formats and various \textit{reduced} formats, while the right plot focuses on the high-purity, high-retention region. Notably, several reduced formats achieve better scores with a large margin compared to the normal format.

\begin{figure}[H]
    \centering
    \includegraphics[width=1\linewidth]{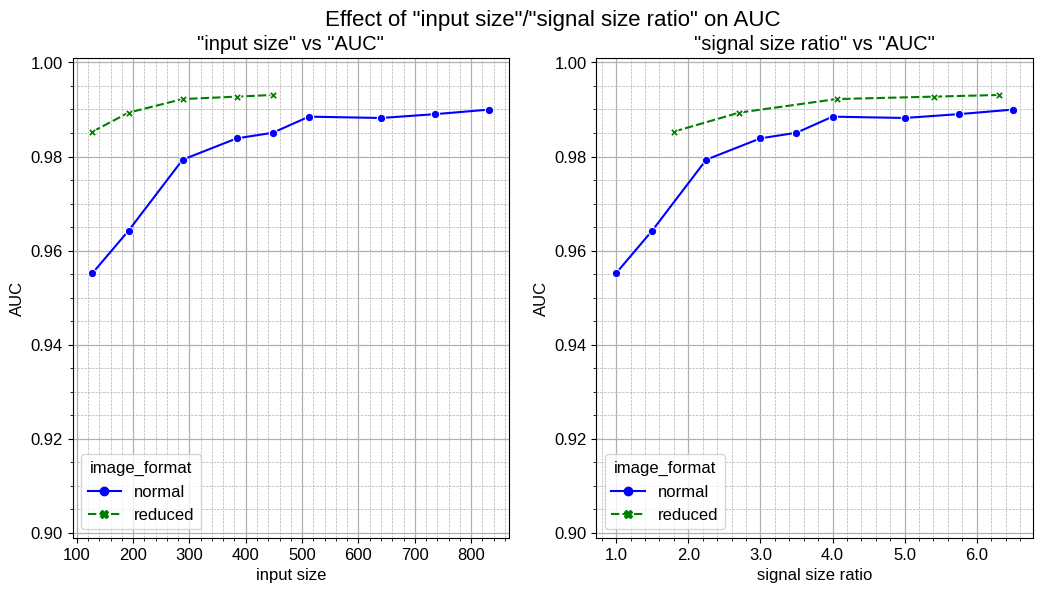}
    \caption{Effect of input size and signal size ratio on test AUC. The left plot shows the relationship between input size and test AUC, while the right plot shows the relationship between signal size ratio and test AUC. The signal size ratio represents the pixel size of  the single cell of the CDC (e.g., 2.0 means one cell occupies two pixels). The signal size ratio is more critical than input size for model performance, as shown by the test AUC (the AUC of the Purity-Retention curve). This explains the superiority of the reduced format over the normal format for Purity-Retention performance. For the reduced format, a signal size ratio of 1 corresponds to an input size of 71, while for the normal format, it corresponds to an input size of 128. The reduced format achieves higher AUC with smaller input sizes, demonstrating its efficiency.}
    \label{fig:test_auc_input_image_size}
\end{figure}

The pattern of signal hit tracks in the normal format is circular and easier to understand for humans, whereas the reduced format stacks the cell of the CDC in a 2D format without any alignment compensation. Therefore, we initially expected the normal format to be better than the reduced format. However, the reduced format's superior performance, as demonstrated in the Purity-Retention curves, can be explained by the signal size ratio shown in Figure \ref{fig:test_auc_input_image_size}.

To delve deeper into the reasons behind these results, we examine the impact of the signal size ratio and input image size, as shown in Figure \ref{fig:test_auc_input_image_size}. The signal size ratio, defined as the pixel size of the single cell of the CDC (e.g., 2.0 means one cell occupies two pixels), is crucial for model performance. This figure highlights two critical insights:

\begin{enumerate}
    \item \textbf{Signal Size Ratio:} The right plot of Figure \ref{fig:test_auc_input_image_size} demonstrates that the signal size ratio is more crucial than the input image size for explaining model performance. A higher signal size ratio corresponds to better model performance, as indicated by the test AUC (Area Under the Curve) of the Purity-Retention curve. This finding explains why the \textit{reduced} format, which achieves a higher signal size ratio with a smaller image size compared to the normal format, outperforms the \textit{normal} format at the same image size.
    \item \textbf{Effect of Initial Layers in Computer Vision Models:} Typical computer vision models, including our backbone model EfficientNet, utilize initial layers that reduce the input image size with a minimal number of layers to lower computational costs. While effective for conventional image classification tasks, this approach can be detrimental in our context, where it is essential to classify scattered or non-consecutive signal hits. The number of parameters for the earlier stages are significantly small, as shown in Table \ref{tab:parameter_counts}. A smaller input size or signal size at these initial layers can result in the loss of fine-grained information such as the original hit location. Consequently, larger input sizes are more beneficial for our task, as they preserve more detailed information on deeper and more extensive layers. However, as the signal size ratio reaches approximately four, which corresponds to the loss starting at the feature map C3, the performance gains plateau, aligning with the expected reduction effect of initial layers.

\end{enumerate}

\begin{table}[h]
\centering
\renewcommand{\arraystretch}{1.0} 
\begin{tabular}{|c|c|c|c|c|c|c|}
\hline
\textbf{Stage} & \textbf{1} & \textbf{2} & \textbf{3} & \textbf{4} & \textbf{5} & \textbf{Total} \\
\hline
\textbf{Feature Map(Size)} & C1 (1/2) & C2 (1/4) & C3 (1/8) & C4 (1/16) & C5 (1/32) & - \\
\hhline{|=======|}
\textbf{Counts [M]} & 0.0047 & 0.029 & 0.11 & 1.4 & 6.2 & \textbf{7.7} \\
\hline
\textbf{Ratio [\%]} & 0.06 & 0.38 & 1.48 & 17.63 & 80.45 & - \\
\hline
\end{tabular}
\caption{Number of parameters and their ratio for different stages in the backbone model. The backbone model is composed of 5 stages, each containing some parameters and producing a feature map. Each stage compresses the feature map size by a factor of 2. The feature map sizes for stages C1, C2, ..., C5 are $1/2$, $1/4$, ..., $1/32$ of the model input size, respectively, as shown in Figure \ref{figure:unet}. This table presents the parameter counts for our default backbone model, EfficientNet-B2.}
\label{tab:parameter_counts}
\end{table}

In summary, our model's ability to achieve the target score is significantly influenced by the use of a reduced format and the careful optimization of input image size. The Purity-Retention curves and the test AUC analysis underscore the importance of these factors in enhancing model performance and achieving the desired outcomes.

\subsection{Effect of Scaling Model Parameters}

\begin{figure}[H]
    \centering
    \includegraphics[width=1.0\linewidth]{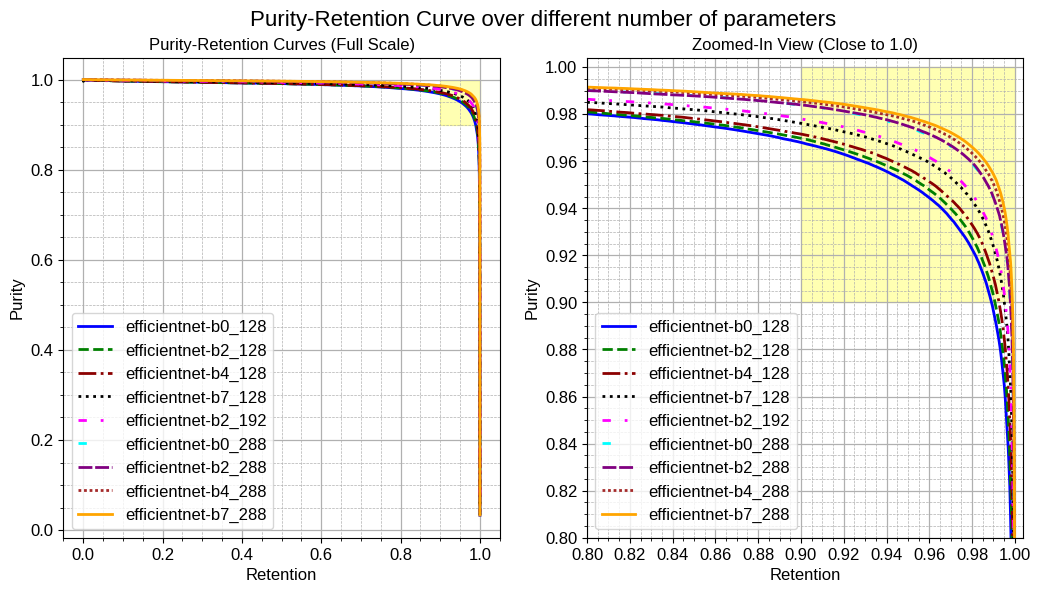}
    \caption{Purity-Retention curves for different EfficientNet models and input image sizes. The left plot shows the overall curves for models, where \textit{EfficientNet-Bx\_yyy} indicates the backbone with EfficientNet-B\textit{x} and its input image size \textit{yyy}, while the right plot zooms in on the high-purity, high-retention region (yellow area). Performance improves with larger models (from B0 to B7) at an input size of 128. However, at an input size of 288, performance is nearly identical across all models, highlighting the more significant impact of image size compared to the number of parameters.}
    \label{fig:pr_curve_effnet_models}
\end{figure}

Our experiments investigated the impact of scaling the number of parameters in our backbone model, EfficientNet, on signal detection performance. We evaluated models ranging from EfficientNet-B0 to EfficientNet-B7, which span approximately 4.0 million to 65 million parameters (Table \ref{tab:efficientnet_parameters}), to determine their efficacy in the context of COMET signal tracking. The results are summarized in Figure \ref{fig:pr_curve_effnet_models}.

\begin{table}[h]
\centering
\begin{tabular}{|l|c|}
\hline
\textbf{Model} & \textbf{Number of Parameters} \\
\hline
EfficientNet-B0 & 4.0M (5.3M) \\
\hline
EfficientNet-B2 (our default) & 7.7M (9.2M) \\
\hline
EfficientNet-B4 & 17.5M (19M) \\
\hline
EfficientNet-B7 & 63.8M (66M) \\
\hline
\end{tabular}
\caption{Number of parameters for different EfficientNet models used as our backbone model. The numbers in parentheses correspond to the original model parameters as reported in \cite{tan2019efficientnet}. Our model does not include the final classification linear layer weights used for the ImageNet benchmark, which accounts for the difference in parameter counts.}
\label{tab:efficientnet_parameters}
\end{table}

Scaling the model from EfficientNet-B0 to EfficientNet-B7 demonstrates a clear performance improvement at an image size of 128. The full-scale and zoomed-in Purity-Retention curves indicate that larger models, such as EfficientNet-B7, achieve higher purity and retention scores compared to smaller models like EfficientNet-B0. This suggests that increasing the number of parameters can enhance the model's ability to detect signals in smaller images. Notably, the performance of EfficientNet-B7 at an image size of 128 is equivalent to EfficientNet-B2 at an image size of 192. On the other hand, when the image size is increased to 288, the performance across all EfficientNet models (B0, B2, B4, and B7) becomes nearly identical. This indicates that the additional information provided by larger images is more critical for model performance than the increased complexity from additional parameters. This aligns with the previous discussion that at a small image size like 128, the loss of fine-grained information on initial layers can be recovered with parameter scaling. However, at a larger image size like 288, the image size is sufficient, and the parameter increase on initial layers does not contribute significantly to the performance.

Another reason for the reduced sensitivity of parameter scaling at larger image sizes is the nature of the current background hits, which are generated with the resampling method (Sec.\ref{sec: bkg_resampling}) and can be considered relatively easy to classify compared to real cases. Consequently, the models can achieve high accuracy even with fewer parameters when the image size is sufficient to capture the necessary details.

To better understand the impact of scaling model parameters, future research will involve using more realistic background hits that are more challenging to classify. These more complex backgrounds are expected to provide a more accurate assessment of the models' capabilities and may reveal more significant performance differences between models with varying numbers of parameters.

In summary, while scaling the number of parameters in EfficientNet models from B0 to B7 shows performance improvements, the contribution of image size is more substantial. At an image size of 288, the performance gains from increasing parameters become very small, highlighting the importance of sufficient image resolution. Future work will focus on using more realistic backgrounds to further evaluate the models' performance.

\subsection{Handling Multiple Hits}
\begin{figure}[H]
\begin{center}
\includegraphics[width=1.0\linewidth]{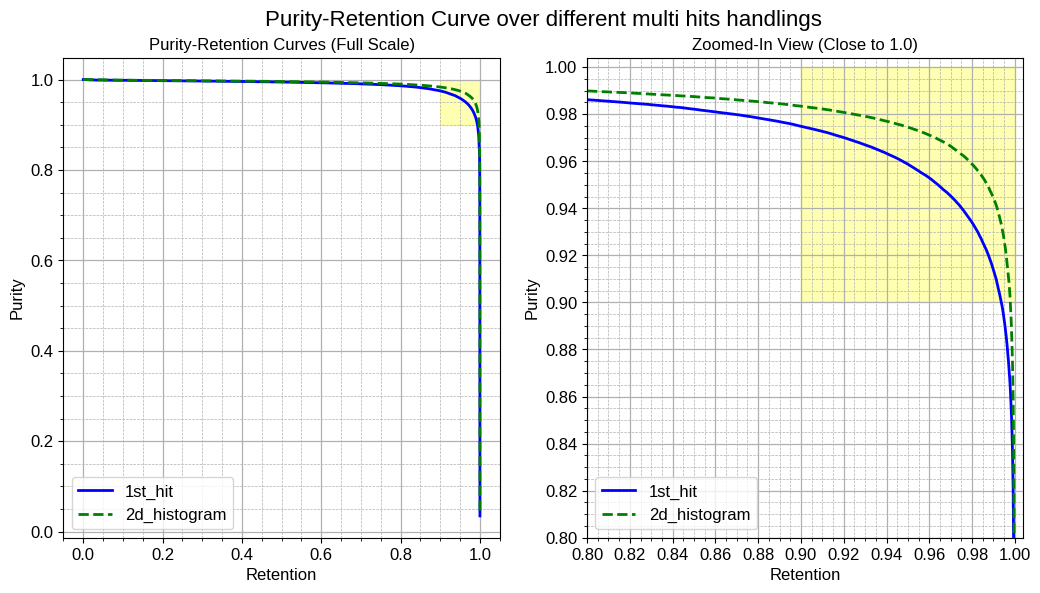}

\caption{Purity-Retention curves for two methods. The left plot shows the full-scale view of the Purity-Retention curves for the \textit{1st\_hit} (solid blue line) and the \textit{highest\_hit} (dashed green line). The \textit{1st\_hit} refers to using the first hit method for multiple hits handling, while the \textit{2D\_histogram} refers to selecting the highest value with the 2D distribution method to improve model performance. The right plot is a zoomed-in view focusing on the high-purity and high-retention region, indicated by the yellow area. This highlighted area represents where both purity and retention are above 0.9, which is our target.}
\label{fig:pr_curve_multi_hits_handling}
\end{center}
\end{figure}

 The Purity-Retention curves in Figure \ref{fig:pr_curve_multi_hits_handling} compare the performance of two preprocessing methods for handling multiple hits within the same cell: the naive 1st-hit method and the 2D distribution  method. The left curve shows the overall Purity-Retention relationship, while the right zooms in on the high-purity region (close to 1.0). The 2D distribution  method (green dashed line) consistently outperforms the 1st-hit method (blue solid line) across almost the entire range of retention values, particularly in the high-retention region, where it maintains higher purity levels compared to the 1st-hit method.

The superior performance of the 2D distribution  method can be attributed to its enhanced signal hit coverage, as highlighted in Table \ref{tab:signal_coverage_comparison}. The table shows that the 1st-hit method has lower overall signal coverage (87\%) and significantly lower signal coverage for multi-hit cells (74\%). In contrast, the 2D distribution  method achieves a higher overall signal coverage (97\%) and better coverage for multi-hit cells (94\%). This improved coverage translates directly into higher purity at comparable retention levels, as demonstrated in the Purity-Retention curves. Although the techniques we explored are relatively basic, their promising results warrant further exploration in future studies.

\begin{table}[h!]
    \centering
    \caption{Comparison of Signal Coverage between the 1st Hit and 2D distribution methods. This statistics is calculated with 8000 events from the training data. The 2D distribution method shows better coverage than the 1st hit method, which explains the superior performance of the 2D distribution in the Purity-Retention curve.}
    \begin{tabular}{lcc}
        \hline
        Method & Overall Signal Coverage & Signal Coverage for Multi-Hit Cells \\
        \hline
        1st Hit & 0.87 (489991 / 560997) & 0.74 (201274 / 272280) \\
        2D distribution  & 0.97 (545466 / 560997) & 0.94 (256749 / 272280) \\
        \hline
    \end{tabular}
    \label{tab:signal_coverage_comparison}
\end{table}

\subsection{Sensitivity to Scaling Methods}

\begin{figure}[H]
\begin{center}
\includegraphics[width=1\linewidth]{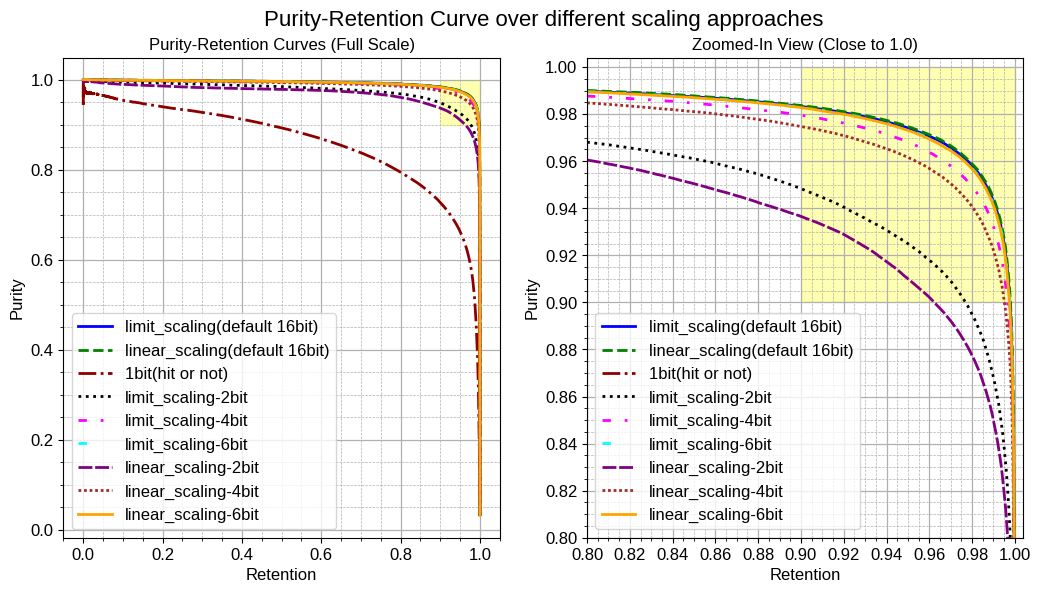}

\caption{Purity-Retention curves over different scaling approaches. The left plot shows the full-scale view of the Purity-Retention curves, while the right plot provides a zoomed-in view focusing on the high-purity and high-retention region (indicated by the yellow area). The curves compare results for \textit{limited\_scaling} (solid lines) and \textit{linear\_scaling} (dashed lines) across multiple bit precision of input: 16-bit (default), 6-bit, 4-bit, 2-bit, and 1-bit, which contains only information about the presence or absence of a hit. The \textit{limited scaling} enhances resolution in signal-dominant areas, while \textit{linear scaling} applies a uniform scale across the range as introduced in Sec.\ref{sec:hits_to_pixs}. This highlighted area represents where both purity and retention are above 0.9, our target performance.}
\label{fig:pr_curve_lim_scaling}
\end{center}
\end{figure}

Another critical aspect of our investigation was the sensitivity of the model to different scaling methods, as illustrated in Figure \ref{fig:pr_curve_lim_scaling}. We compared two scaling techniques: the limited scaling and the linear scaling. The limited scaling method was designed to double the resolution for critical ranges of hit timing and ADC compared to the linear scaling as discussed in Sec.\ref{sec:hits_to_pixs}.

In the default precision of 16-bit, the performance metrics were nearly identical for both scaling methods, indicating that the linear scaling method already provides sufficient resolution for our purposes. This outcome suggests the robustness of our model across different scaling approaches.

\begin{figure}[H]
    \begin{minipage}{0.49\linewidth}
        \centering
        \includegraphics[width=\linewidth]{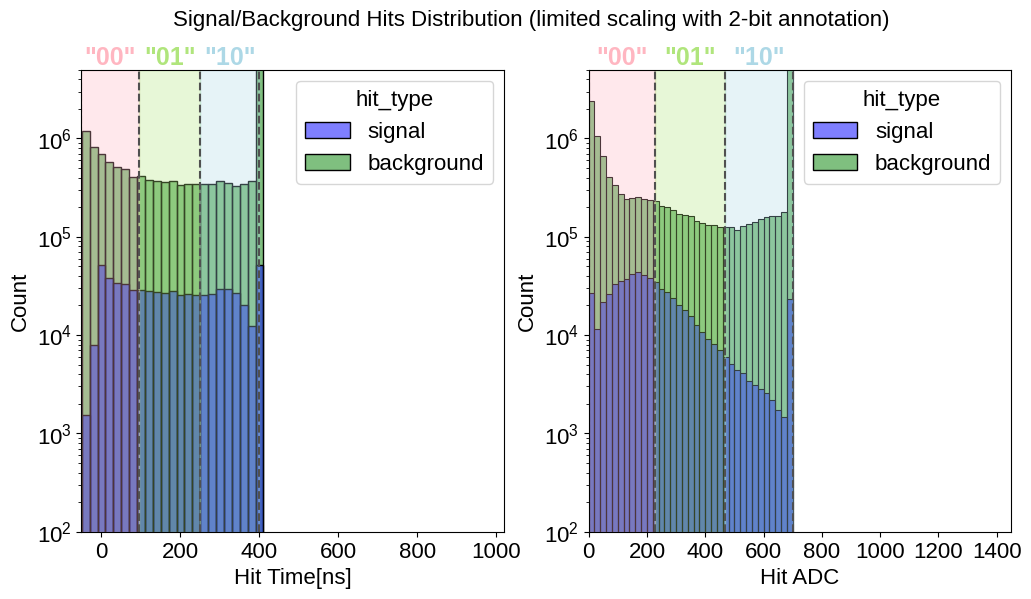}
        \caption*{(a)}
    \end{minipage}
    \hfill
    \begin{minipage}{0.49\linewidth}
        \centering
        \includegraphics[width=\linewidth]{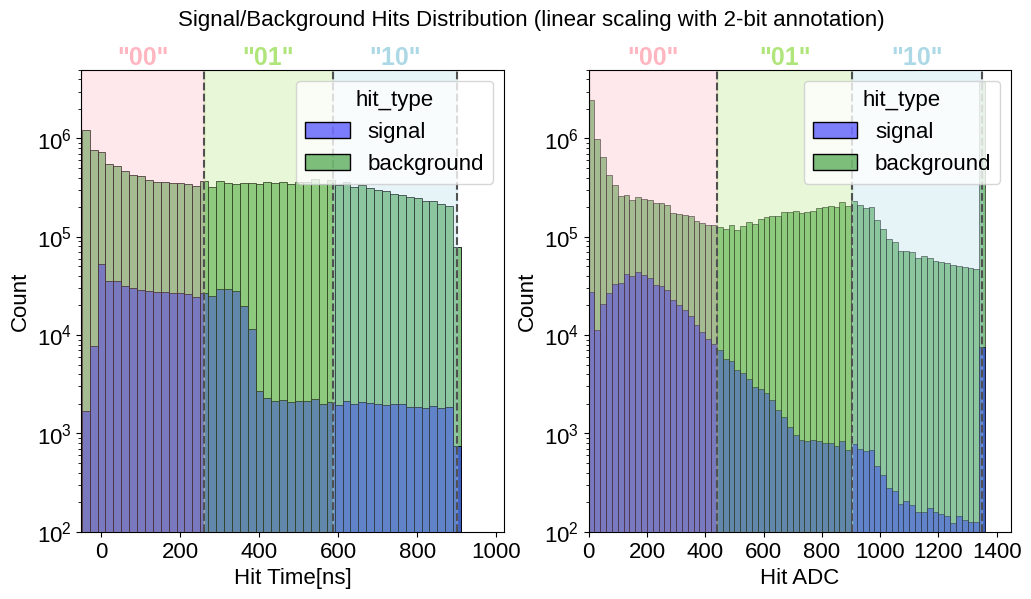}
        \caption*{(b)}
    \end{minipage}
    \caption{Comparison of signal and background hit distributions with the annotation of 2-bit input case: (a) using limited scaling and (b) using linear scaling. Both plots show the distribution of hits over scaled hit timing and ADC with a log-scaled y-axis. The shaded areas represent different states for low-bit precision inputs, specifically in the case of 2-bit precision. Only three states ("00", "01", "10") are illustrated in the graphs; the "11" state is reserved for no hit and is not displayed here. The limited scaling method in (a) provides finer resolution in signal-dominant areas, enhancing model performance in low-bit precision scenarios.}
    \label{fig:signal_bkg_distribution_with_2bit_annot}
\end{figure}

To further investigate, we analyzed the performance with low-bit precision inputs. As expected, with lower precision such as 2-bit and 4-bit, the limited scaling method outperformed the linear scaling. As illustrated in Figure \ref{fig:signal_bkg_distribution_with_2bit_annot}, the 2-bit case shows that the limited scaling results in finer resolution in the critical ranges of hit timing and ADC, highlighting its effectiveness in retaining important features in low-bit scenarios.

Interestingly, even at 6-bit precision, the performance metrics remained nearly identical to both the 16-bit results, demonstrating minimal sensitivity to bit precision at this level. This result implies that the linear scaling might be preferred as it covers the entire range of input data more uniformly because we can use higher bit data like 16-bit.

Surprisingly, even the lowest precision of 1-bit demonstrated some level of performance, which means the model can predict correctly based on only the hit location. This finding suggests that spatial information alone can provide valuable insight for the model.

The moderate sensitivity to scaling methods, particularly at higher bit precision, positively indicates the robustness of our model. This suggests its suitability for a range of input conditions without requiring complex scaling adjustments. Thus, we can confidently use the simple linear scaling method in subsequent studies without compromising performance.

\subsection{Sensitivity to $\Delta\phi$ Feature}

\begin{figure}[H]
    \centering
    \includegraphics[width=1.0\linewidth]{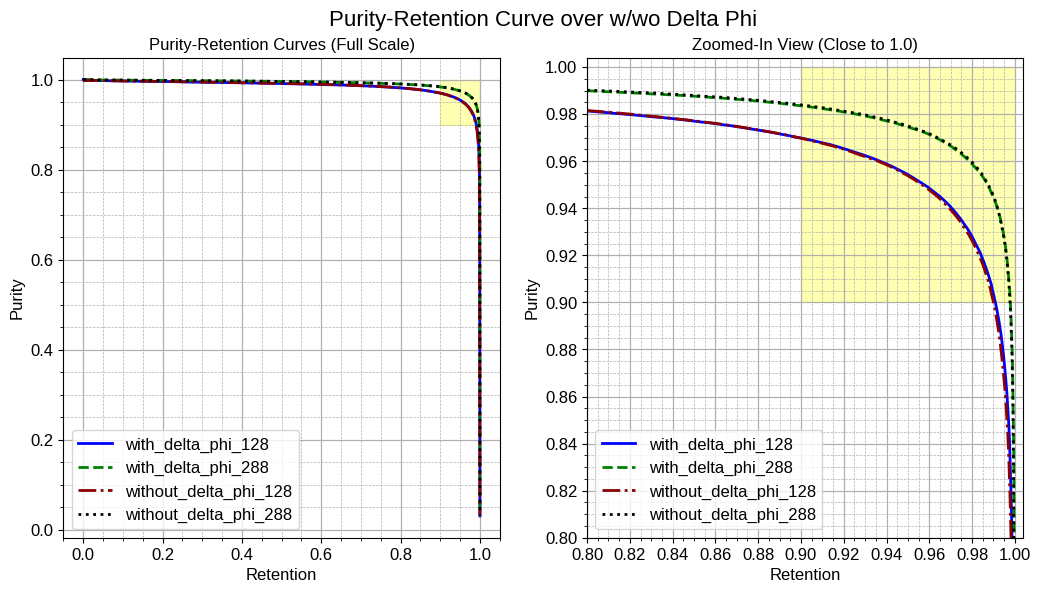}
    \caption{Purity-Retention curves for models trained with and without the $\Delta\phi$ feature. The left plot shows the full-scale view of the curves for different input sizes, with \textit{with\_delta\_phi\_xxx} and \textit{without\_delta\_phi\_xxx} indicating the presence or absence of the $\Delta\phi$ feature for input size \textit{xxx}. The right plot zooms in on the high-purity, high-retention region (yellow area), where both metrics exceed 0.9. The results indicate that the inclusion of the $\Delta\phi$ feature does not significantly affect the model performance for both input sizes.}
    \label{fig:pr_curve_delta_phi}
\end{figure}

To assess the impact of including the $\Delta\phi$ feature on model performance, we conducted experiments comparing models trained with and without this feature, which corresponds to 3-channel and 2-channel input images in our setup respectively. The $\Delta\phi$ feature represents the azimuthal angle difference between the CTH trigger counter and each cell, which could potentially enhance the model's ability to distinguish signal hits from background hits.
Our evaluation focused on two primary input image sizes: 128$\times$128 and 288$\times$288. The Purity-Retention curves for these configurations are presented in Figure \ref{fig:pr_curve_delta_phi}.

The results show that the models trained with and without the $\Delta\phi$ feature exhibit similar performance in terms of Purity and Retention across both input sizes, with slightly better performance when including the $\Delta\phi$ feature. This suggests that the hit timing and ADC features are sufficient to classify the signal and background hits effectively, and the addition of the $\Delta\phi$ feature does not provide significant additional discriminatory power.

Based on the sensitivity analysis, we conclude that the $\Delta\phi$ feature is not essential for achieving high performance in the current setup. Future work may involve exploring other features to further enhance model performance. Additionally, evaluating the $\Delta\phi$ feature with more realistic background hits could provide further insights into its potential utility.

\section{Conclusion}
This paper has presented a pioneering approach to tracking analysis within the COMET experiment, utilizing deep learning techniques of semantic segmentation. Our conclusions are as follows:

\begin{enumerate}
    \item \textbf{Achievement of Target Performance Goals.} Our model achieved a purity rate of 98\% and a retention rate of 90\%, surpassing our target performance definitions of a minimum purity and retention of 90\%. This high level of accuracy, attained using standard semantic segmentation methods without complex adjustments, highlights the potential of deep learning in enhancing precision in the COMET analysis. Furthermore, this study represents the initial application of deep learning methods to the COMET tracking. While the methods used were relatively straightforward, they have opened up possibilities for more advanced techniques promising even more refined solutions in future research.

    \item \textbf{Critical Factors in Model Performance:}
    \begin{itemize}
        \item \textbf{Input Image Size and Signal Size Ratio:} Larger input image sizes help preserve more fine-grained information, which is considered to be crucial for accurately classifying scattered or non-consecutive signal hits. The signal size ratio, defined as the pixel size of the single cell of the CDC, was found to be more critical than the input image size for model performance. Higher signal size ratios correspond to better performance, explaining why the reduced format, which achieves higher signal size ratios, outperforms the normal format. However, the performance gains plateau as the signal size ratio reaches approximately four. This trend aligns with the parameter allocation of the backbone model, which means fewer parameters in the initial layers.
        \item \textbf{Handling Multiple Hits:} Our study identified the handling of multiple hits on the same CDC cell as a crucial factor. Our current approach yielded satisfactory results, but we anticipate that more tailored methods could further improve model performance.
        \item \textbf{Sensitivity to Scaling Methods:} Contrary to initial concerns, the model showed robust performance across different scaling techniques. This suggests that the final performance is not overly sensitive to scaling adjustments, thus reducing the need for complex and precise scaling configurations.
    \end{itemize}

  \item \textbf{Broader Context and Future Directions:} While this paper focuses on signal track extraction 
  in the COMET experiment, it represents only a part of the broader range of the COMET analysis. The ultimate goal of the COMET experiment extends beyond tracking, aiming to determine the momentum of electrons from signal hits. Our future work will continue to explore and refine deep learning approaches to meet this objective, including choosing appropriate backbone models and other hyperparameter tuning. Initially, we will focus on developing robust models with simulation data by incorporating more realistic background hits. Following this, we will address the gap between simulation and real data to ensure our methods are robust and reliable in practical applications.
\end{enumerate}

In conclusion, this study demonstrates not only the feasibility but also the effectiveness of deep learning in addressing complex tracking challenges in the COMET experiment. The encouraging results of this research pave the way for further exploration and application of advanced deep learning techniques in this field, potentially significantly enhancing our understanding and analysis capabilities in high-energy physics experiments.

\section{Acknowledgement}
This work was supported by the Japan Society for
the Promotion of Science (JSPS) KAKENHI, under Grant Numbers JP18H05231 (YK),
JP18H01210 and JP18H05543 (JS).

\bibliographystyle{elsarticle-num}
\bibliography{references}

\end{document}